\documentclass[conference, letterpaper, 10pt, final, comsoc]{IEEEtran}

\IEEEoverridecommandlockouts

\usepackage[latin1]{inputenc}
\usepackage{chngcntr}
\usepackage[cmex10]{amsmath}
\usepackage[T1]{fontenc}
\usepackage{amsfonts}
\usepackage{amssymb}
\usepackage{graphicx}
\usepackage{mathtools}
\usepackage{bbm}
\usepackage[normalem]{ulem}
\usepackage{tabularx}
\usepackage[nottoc]{tocbibind}
\usepackage{float}
\usepackage{enumerate}
\usepackage[numbers,sort&compress]{natbib}
\usepackage[dvipsnames]{xcolor}
\usepackage{tikz}
\usepackage{physics}
\usepackage{todonotes}
\usepackage{algorithm}
\usepackage[noend]{algpseudocode}
\usepackage[final]{microtype}
\usepackage{dsfont}
 \usepackage{hyperref}
\usetikzlibrary{arrows.meta}

\usepackage{algorithm}
\usepackage{algpseudocode}
\algnewcommand\algorithmicforeach{\textbf{for each}}
\algdef{S}[FOR]{ForEach}[1]{\algorithmicforeach\ #1\ \algorithmicdo}

\tikzset{>=latex}

\algnewcommand{\LeftComment}[1]{\Statex \(\triangleright\) #1}

\makeatletter
\renewcommand{\ALG@name}{Protocol}
\renewcommand{\complement}{{\mathsf{c}}}
\makeatother
\algblock{Input}{EndInput}
\algnotext{EndInput}
\algblock{Output}{EndOutput}
\algnotext{EndOutput}
\algnotext{EndIf}
\algrenewcommand{\algorithmicthen}{:}

\usetikzlibrary{arrows,
                positioning,
                decorations.pathmorphing,
                decorations.markings,
                decorations.pathreplacing,
                shapes,
                fadings,
                calc
            }
\counterwithin{equation}{section}
\newcounter{thm}
\newtheorem{theorem}[thm]{Theorem}

\newtheorem{remark}[thm]{Remark}
\newtheorem{definition}[thm]{Definition}

\newtheorem{lemma}[thm]{Lemma}

\DeclareMathOperator{\efd}{efd}
\newcommand*{\rom}[1]{\uppercase\expandafter{\romannumeral #1\relax}}

\newcommand{\eps}{{\varepsilon}}        %%%%%%%%%%%%%%%%%%%%%%%%%%%%%%%%%%%

\newcommand{\eins}{{\mathbbm{1}}}

\allowdisplaybreaks

\makeatletter
\newcommand*{\algrule}[1][\algorithmicindent]{%
  \makebox[#1][l]{%
    \hspace*{.2em}% <------------- This is where the rule starts from
    \vrule height .75\baselineskip depth .25\baselineskip
  }
}

\newcount\ALG@printindent@tempcnta
\def\ALG@printindent{%
    \ifnum \theALG@nested>0% is there anything to print
    \ifx\ALG@text\ALG@x@notext% is this an end group without any text?
    \else
    \unskip
    \ALG@printindent@tempcnta=1
    \loop
    \algrule[\csname ALG@ind@\the\ALG@printindent@tempcnta\endcsname]%
    \advance \ALG@printindent@tempcnta 1
    \ifnum \ALG@printindent@tempcnta<\numexpr\theALG@nested+1\relax
    \repeat
    \fi
    \fi
}

\newcounter{protocol}

\counterwithout{equation}{section} % Remove the resetting by `section`

\title{Compound Channel Capacities under Energy Constraints and Application}
\author{
    \IEEEauthorblockN{Andrea Cacioppo\IEEEauthorrefmark{1}, Janis N\"otzel\IEEEauthorrefmark{1}}
    \IEEEauthorblockA{
        \IEEEauthorrefmark{1}Emmy-Noether Gruppe Theoretisches Quantensystemdesign\\Lehrstuhl f\"ur Theoretische Informationstechnik \\
        Technische Universit\"at M\"unchen\\
        \{andrea.cacioppo,janis.noetzel\}@tum.de
    }
    \and 
    \IEEEauthorblockN{Matteo Rosati\IEEEauthorrefmark{2}}
    \IEEEauthorblockA{
            \IEEEauthorrefmark{2}
             Departament de F\'{\i}sica: Grup d'Informaci\'{o} Qu\`{a}ntica,\\
             Universitat Aut\`{o}noma de Barcelona,\\
             ES-08193 Bellaterra (Barcelona), Spain.\\
            matteo.rosati@uab.cat
    }
}

\begin{document}

\maketitle
\begin{abstract}
    Compound channel models offer a simple and straightforward way of analyzing the stability of decoder design under model variations. With this work we provide a coding theorem for a large class of practically relevant compound channel models. We give explicit formulas for the cases of the Gaussian classical-quantum compound channels with unknown noise, unknown phase and unknown attenuation. We show analytically how the classical compound channel capacity formula motivates nontrivial choices of the displacement parameter of the Kennedy receiver. Our work demonstrates the value of the compound channel model as a method for the design of receivers in quantum communication.
    {\let\thefootnote\relax\footnote{
\copyright 2020 IEEE. Personal use of this material is permitted. Permission from IEEE must be obtained for all other uses, in any current or future media, including reprinting/republishing this material for advertising or promotional purposes, creating new collective works, for resale or redistribution to servers or lists, or reuse of any copyrighted component of this work in other works.}}

\end{abstract}
%%%%%%%%%%%%%%%%%%%%%%%%%%%%%%%%%%%%%%%%%%%%%%%%%%%%%%%%%%%%%%%%%%%%%%%%%
%%%%%%%%%%%%%%%%%%%%%%%%%%%%%%%%%%%%%%%%%%%%%%%%%%%%%%%%%%%%%%%%%%%%%%%%
\section{Introduction}
Compound channels model transmission over a noisy communication line when the noise level is not known prior to transmission, but rather only guaranteed to lie within some region. This communication model is thus closer to application than the typical i.i.d. model. A typical strategy to resolve the uncertainty in this setting is for the sender to transmit pilot symbols, in which case the receiver is able to estimate the channel parameter. However, this strategy only affects the capacity in case that the sender knows the exact noise level in advance, or else there is a feedback loop from sender to receiver. While this lets the model appear as a suitable tool for optimization of real-world communication systems, the current literature on compound quantum channel has so far not considered infinite-dimensional systems, and this omission led to a lack of applicability of the model. With this work we take a first step to closing this gap, by providing several explicit capacity formulas for classical-quantum channels with unknown Gaussian noise, unknown phase shift, and unknown attenuation level, which model typical noise effects in fiber-optical and free-space communication \cite{attenuation-qkd-npj-paper,photonic-network-paper,Waseda:11}.

Moreover, we apply the theory of \emph{classical} compound channels to the optimization of a Kennedy receiver when applied to a classical-quantum compound attenuation channel, thereby promoting the application of the theory to receiver design for quantum communication systems.

%%%%%%%%%%%%%%%%%%%%%%%%%%%%%%%%%%%%%%%%%%%%%%%%%%%%%%%%%%%%%%%%%%%%%%%%%
%%%%%%%%%%%%%%%%%%%%%%%%%%%%%%%%%%%%%%%%%%%%%%%%%%%%%%%%%%%%%%%%%%%%%%%%%
%\red{
\subsection{RELATED WORK}
The study of compound channels can be traced back to the work of Blackwell, Breiman and Thomasian \cite{bbt59}. A full coding theorem for finite-dimensional classical-quantum compound channels was obtained independently in \cite{bb2009} and in \cite{hayashi2009}. The technical foundations of this work are the direct coding theorem as proven in \cite{bbjn2012}, the converse for the averaged channel as in \cite{datta-dorlas2007}, and the approximation tools as presented in \cite{winter2016tight}, which have already been applied successfully for the derivation of coding theorems for memoryless channels in \cite{shirokov-uniform-approximation}.  

%The first compound channel coding theorems in the domain of quantum communication have been derived already in 2009 \cite{bb2009, hayashi2009} and 
%... there is one strategy known which applies to finite averaged channels and should work for compound channels as well. It uses pilot symbols to let the receiver detect its ``branch'' of the channel, while the sender uses a code-book that works for every branch. Strictly speaking, it does not apply here.
%\cite{shirokov2006}}
%%%%%%%%%%%%%%%%%%%%%%%%%%%%%%%%%%%%%%%%%%%%%%%%%%%%%%%%%%%%%%%%%%%%%%%%%
%%%%%%%%%%%%%%%%%%%%%%%%%%%%%%%%%%%%%%%%%%%%%%%%%%%%%%%%%%%%%%%%%%%%%%%%%
%%%%%%%%%%%%%%%%%%%%%%%%%%%%%%%%%%%%%%%%%%%%%%%%%%%%%%%%%%%%%%%%%%%%%%%%%
%%%%%%%%%%%%%%%%%%%%%%%%%%%%%%%%%%%%%%%%%%%%%%%%%%%%%%%%%%%%%%%%%%%%%%%%%
\section{Notation}
The set indexing the signal states is written as $\mathbf X$ if finite and as $\mathcal X$ if infinite. Likewise, $\mathbf S$ and $\mathcal S$ denote finite and infinite sets of channel states. Hilbert spaces are denoted as $\mathcal H$, their dimensions as $\mathrm{dim}\mathcal H$. Throughout, they are assumed to be separable. The set of probability measures on a set $\mathcal A$ is written $\mathcal P(\mathcal A)$, the set of states on $\mathcal H$ is $\mathcal P(\mathcal H)$. We denote by $\mathcal P_f(\mathcal A)$ the set of distributions with $p(a)>0$ for a finite set of points. The trace of an operator $A$ on $\mathcal H$ is denoted $tr(A)$, the scalar product of $x,y\in\mathcal H$ as $\langle x,y\rangle$. The logarithm $\log$ is taken with respect to base two, and the entropy of $\rho\in\mathcal P(\mathcal H)$ or $p\in\mathcal P(\mathcal A)$ is then written as $H(\rho):=-tr(\rho\log\rho)$ or $H(p):=-\sum_{a\in\mathcal A}p(a)\log p(a)$. The binary entropy is $h:[0,1]\to[0,1]$. Classical-quantum channels will be denoted as $\mathcal N$. The set of all such channels, with input set $\mathcal X$ and output space $\mathcal H$, is $C(\mathcal X,\mathcal H)$. The set of all classical channels with given input alphabet $\mathbf X$ and output alphabet $\mathbf Y$ is denoted $C(\mathbf X,\mathbf Y)$. The Holevo quantity of a distribution $p\in\mathcal P(\mathcal X)$ and classical-quantum channel $\mathcal N\in C(\mathcal X,\mathcal H)$ is $\chi(p,\mathcal N)=H(\int p(x)\mathcal N(x)dx)-\int p(x)H(\mathcal N(x))dx$. The one-norm is denoted as $\|\cdot\|_1$. For our analysis of the Dolinar receiver we make use of the Frobenius norm $\|\cdot\|_F$. For $a\in[0,1]$ we abbreviate $1-a$ as $a'$.
For a Hamiltonian $\mathbb H$ on $\mathcal H$, and $E\in\mathbb R$, we let $\mathcal P_{\mathbb H,E}:=\{\rho:tr(\rho\mathbb H)\leq E\}$. The relative entropy of $\rho,\sigma\in\mathcal P(\mathcal H)$ is $D(\rho\|\sigma)$.

%%%%%%%%%%%%%%%%%%%%%%%%%%%%%%%%%%%%%%%%%%%%%%%%%%%%%%%%%%%%%%%%%%%%%%%%%
%%%%%%%%%%%%%%%%%%%%%%%%%%%%%%%%%%%%%%%%%%%%%%%%%%%%%%%%%%%%%%%%%%%%%%%%%
%%%%%%%%%%%%%%%%%%%%%%%%%%%%%%%%%%%%%%%%%%%%%%%%%%%%%%%%%%%%%%%%%%%%%%%%%
%%%%%%%%%%%%%%%%%%%%%%%%%%%%%%%%%%%%%%%%%%%%%%%%%%%%%%%%%%%%%%%%%%%%%%%%%
\section{Definitions}
A classical-quantum compound channel is any set $\mathcal N=\{\mathcal N_s\}_{s\in\mathcal S}$ of channels where $\mathcal N_s\in C(\mathcal X,\mathcal H)$ for every $s\in\mathcal S$. 
\begin{definition}[$(n,\lambda)$ Code] 
    An $(n,\lambda)$ code $\mathcal C$ for the compound channel $\mathcal N$ consists of a finite collection $\{x_m\}_{m=1}^M\subset\mathcal X^n$ of signals and a POVM $\{D_m\}_{m=1}^M$. If the success probability 
    \begin{align}
        p_{\mathrm{success}}(\mathcal C):=\inf_{s\in\mathcal S}\frac{1}{M}\sum_{m=1}^Mtr(D_m \mathcal N_s^{\otimes n}(x_m))
    \end{align}
    of $\mathcal C$ satisfies $p_s(\mathcal C)\geq1-\lambda$, it is called an $(n,\lambda)$ code. If each $x_i$ is constrained to lie inside a set $\mathcal R\subset\mathcal X$, then $\mathcal C$ is said to obey state constraint $\mathcal R$.
\end{definition}

\begin{definition}[Achievable Rates, Capacity]
    A rate $R\geq0$ is called achievable for the classical-quantum compound channel $\mathcal N$ under state constraint $\mathcal R$ if there exists a sequence $(\mathcal C_n)_{n\in\mathbb N}$ of $(n,\lambda_n)$  codes, obeying the state constraint $\mathcal R$, such that both $\lambda_n\to0$ and $\limsup_{n\to\infty}\frac{1}{n}\log M_n\geq R$. 
    
    The message transmission capacity of $\mathcal N$ under average error criterion is defined as the supremum over all rates that are achievable for $\mathcal N$. It is denoted as $C(\mathcal N)$ here for brevity.
\end{definition}
The following definition is an important ingredient to our analysis, which depends to a large degree on tools developed for finite-dimensional quantum channels:
\begin{definition}[Effective Dimension]
Let $\eps>0$. A compound channel $\mathcal N$ is said to have effective dimension $\efd(\epsilon)\in \mathbb{N}$ with respect to a constraint $\mathcal R\subset\mathcal X$ if there exists a projector P s.t.
\begin{align}
      \forall s\in\mathcal S,x\in\mathcal R:\  &tr(P \mathcal N_s(x))   \geq 1 - \epsilon\  \wedge  \ tr(P)   \leq \efd(\eps).
\end{align}

\end{definition}
%%%%%%%%%%%%%%%%%%%%%%%%%%%%%%%%%%%%%%%%%%%%%%%%%%%%%%%%%%%%%%%%%%%%%%%%%
%%%%%%%%%%%%%%%%%%%%%%%%%%%%%%%%%%%%%%%%%%%%%%%%%%%%%%%%%%%%%%%%%%%%%%%%%
%%%%%%%%%%%%%%%%%%%%%%%%%%%%%%%%%%%%%%%%%%%%%%%%%%%%%%%%%%%%%%%%%%%%%%%%%
%%%%%%%%%%%%%%%%%%%%%%%%%%%%%%%%%%%%%%%%%%%%%%%%%%%%%%%%%%%%%%%%%%%%%%%%%
\section{Results}
Before listing our results, we cite here one of our main technical tools, which asserts the continuity of the entropy on energy shells $\mathcal P_{\mathbb H,E}$. Since our main focus is the derivation of capacity formulas for models of potential practical interest, we can assume our communication systems as equipped with a Hamiltonian $\mathbb{H}$ describing the dynamics of the output system. 
%Mathematically, we then use $\mathbb{H}$ to describe the average energy of the output system. 
Throughout, we will assume that the Hamiltonian obeys the Gibbs hypothesis, which, for brevity, we note down here together with an important consequence:  
\begin{lemma}[{\cite[Gibbs Hypothesis; Lemma 15]{winter2016tight}}]\label{lem:gibbs-hypothesis}
    If $tr\left(\exp{-\beta \mathbb{H}}\right)<\infty$ for all $\beta>0$ and $\max\{tr(\rho \mathbb{H}),tr(\sigma \mathbb{H})\}\leq E$ for some $E>0$ then $\|\rho-\sigma\|_1\leq\eps$ implies $|H(\rho) - H(\sigma)|<\eps C(\eps,\mathbb{H},E) + h(\eps)$ for a function $C$ satisfying $\lim_{\eps\to0}\eps C(\eps,\mathbb{H},E)=0$.
\end{lemma}
If $\mathbb H$ satisfies the Gibbs hypothesis, a multitude of techniques for finite-dimensional systems carries over \cite{shirokov-uniform-approximation}, which lets us prove the following statement:

\begin{theorem}\label{thm:compound-coding-theorem}
Let $\mathcal N:=\{\mathcal N_s\}_{s\in\mathcal S}\subset C(\mathcal X,\mathcal H)$ be a compound channel. Let the constraint $x\in \mathcal R$ be imposed on all signals $x$, where $\mathcal R\subset\mathcal X$ such that
$\efd(k^{-2})\in\mathcal O(k^{2^{-1}-\eps})$ for some $\eps>0$.  
%\red{this implies  $\lim_{k\to\infty}\tfrac{\efd(k^{-2})^2}{k}\log(k)=0$ and $\lim_{k\to\infty}(k^{\efd(k^{-2})}2^{-ck}=0$}. 
Let $\sup_{s\in\mathcal S}\sup_{x\in\mathcal R}tr(\mathbb{H}\mathcal N_{s}(x))\leq E_\mathrm{out}$ for some $E_\mathrm{out}\in\mathbb R$. The capacity of $\mathcal N$ is given by 
\begin{align}
    C(\mathcal N)=\sup_{p\in\mathcal P_{\mathcal R}(\mathcal X)}\inf_{s\in\mathcal S}\chi(p;\mathcal N_s)
\end{align}
where $\mathcal P_\mathcal{R}(\mathcal X):=\{p\in\mathcal P_f(\mathcal X):A\cap\mathcal R=\emptyset\Rightarrow p(A)=0\}$.
\end{theorem}
\begin{remark}
    In our examples, we will use the Hamiltonian $\mathbb H=\sum_{n=0}^\infty n\cdot|n\rangle\langle n|$, where $|n\rangle$ are the photon number states.
\end{remark}

We can apply the above results to a variety of channels of practical interest:
\begin{theorem}[Unknown Gaussian Noise]\label{thm:unknown-noise}
    Let $\mathcal X=\mathbb C$ and $\mathcal S=\{\sigma:\sigma\in[A,B]\}$ for some $0\leq A\leq B$. Let $\mathcal N:=\{\mathcal N_\sigma\}_{\sigma\in\mathbf S}$ be a compound channel, where for each $\sigma$
    \begin{align}\label{eq:AddNoise}
        \mathcal N_\sigma(\alpha)=\frac{1}{\sigma\pi}\int\exp{-\frac{|z-\alpha|^2}{\sigma}}|z\rangle\langle z|dz
    \end{align} 
    is a Gaussian channel as in \cite[equation (82)]{holevo1979}. Let $\mathcal R:=\{\alpha:|\alpha|^2\leq E\}$ be the energy constraint on the input states for some $E>0$. The capacity of $\mathcal N$ is then given by 
    \begin{align}
        C(\mathcal N)=g(B+E)-g(B).
    \end{align}
\end{theorem}
\begin{theorem}[Unknown Phase]\label{thm:unknown-phase}
    Let $\mathcal X=\mathbb C$ and $\mathcal S=\{\exp{\mathbbm i\theta}:\theta\in[0,2\pi)\}$. Let for some $\sigma>0$
    \begin{align}
        \mathcal N_s(\alpha)=\frac{1}{\sigma\pi}\int\exp{-\frac{|z-s\alpha|^2}{\sigma}}|z\rangle\langle z|dz.
    \end{align} 
    The capacity of $\mathcal N$ is given by 
    \begin{align}
        C(\mathcal N)=g(\sigma+E)-g(\sigma).
    \end{align}
\end{theorem}
\begin{theorem}[Unknown Attenuation]\label{thm:unknown-attenuation}
    Let $\mathcal X=\mathbb C$ and $\mathcal S=[A,B]$ with $A\geq0$. Let for some $\sigma>0$
    \begin{align}
        \mathcal N_s(\alpha)=\frac{1}{\sigma\pi}\int\exp{-\frac{|z-\sqrt{s}\alpha|^2}{\sigma}}|z\rangle\langle z|dz,
    \end{align} 
    or for $\sigma=0$, $\mathcal N_s(\alpha)=|\sqrt{s}\alpha\rangle$. The capacity of $\mathcal N$ is given by 
    \begin{align}
        C(\mathcal N)=g(\sigma+A\cdot E)-g(\sigma),
    \end{align}
\end{theorem}

Our approach to proving these statements rests on two pillars. First, we employ the proof of the classical-quantum compound channel coding theorem as in \cite{bbjn2012}, which gives error bounds that do not depend on any particular state. 

The proofs in \cite{bbjn2012} make use of the method of types, and thereby the dimension of the involved systems enters in the form of estimates using e.g. that $\tfrac{d^2\log(k)}{k}\rightarrow0$ as $k\to\infty$. Using our requirements on the effective system dimensions $d(k)$, we are able to guarantee, in such cases, that e.g. $\lim_{k\to\infty}\tfrac{d(k)^2\log(k)}{k}=0$. The corresponding proofs can be found in the Appendix. As our examples in Theorems \ref{thm:unknown-noise} - \ref{thm:unknown-attenuation} show, the requirements are satisfied in many situations of potential practical interest. The second important ingredient is the continuity of the entropy on the sets $\mathcal P_{\mathbb H,E}$ \cite{wehrl1978} with respect to the trace norm, in the explicit form as given in \cite{winter2016tight} which, in technical terms, is the replacement of the Fannes-Audenaert inequality \cite{fannes,audenaert2006sharp}.

To prove the explicit formulas, we require corresponding bounds on the effective dimensions. A straightforward way of getting an idea of the effective dimensions for a Gaussian system is to look at effective dimensions needed to cover the statistics of Gaussian states:
\begin{lemma}\label{lem:effectively-finite-dimensionanl-projection}
    There is a sequence $(P_N)_{N\in\mathbb N}$ of projectors such that for every coherent state $|\alpha\rangle$ it holds 
    \begin{align}
        tr(P_N|\alpha\rangle\langle\alpha|)&\geq 1 - 2\exp(-|\alpha|^2)\frac{|\alpha|^{2N}}{N!}\\
        tr(P_N)&=N
    \end{align}
\end{lemma}
Motivated by this promising estimate, we then proceed to prove a tail bound for a Gaussian distribution on the complex plane:
\begin{lemma}\label{lem:energy-bound-for-gaussian-noise}
    Let $C_{E'}:=\{z:|z|^2\leq E'\}^\complement$ and $|\alpha|^2\leq E$. For every $E'>0$ we have
    \begin{align}
        \frac{1}{\pi \sigma}\int_{C_{E+E'}}\exp{-\tfrac{|\alpha-z|^2}{\sigma}}dz\leq e^{-\frac{E'}{\sigma}}
    \end{align}
    Where $^\complement$ indicates the complementary set. In particular, the probability of finding a coherent state $z$ with $|z|^2>E+E'$ at the output of channel \eqref{eq:AddNoise}, upon input of a coherent state $\alpha$ with $|\alpha|^2\leq E$, is upper bounded by $\exp(-E'/\sigma)$.
\end{lemma} 
\begin{remark}\label{rem:application-of-stirling}
    Using the version $\sqrt{2\pi}exp{\{-n\}}n^{n+\tfrac{1}{2}}\leq n!$ of Stirling's formula, the inequality $e^{-x}<1$ (if $x>0$), the estimate $2<\sqrt{6}<\sqrt{2\pi}$ and the assumption $|\alpha|^2\leq E$, we can transform the lower bound on $tr(P_N|\alpha\rangle\langle\alpha|)$ into
    \begin{align}
        tr(P_N|\alpha\rangle\langle\alpha|)
        %JN: I have hidden the detailed steps below
        %\\tr(&P_N|\alpha\rangle\langle\alpha|)\\
            %&\geq1-\tfrac{2}{\sqrt{2\pi}}\exp(-|\alpha|^2)|\alpha|^{2(N+1)}\exp{N}N^{-N-\tfrac{1}{2}}\\
            %&\geq1-\tfrac{2}{\sqrt{2\pi}}|\alpha|^{2(N+1)}\exp{N}\exp{-(N+\tfrac{1}{2})\log N}\\
            %&\geq1-|\alpha|^{2(N+1)}\exp{N}\exp{-(N+\tfrac{1}{2})\log N}\\
            %&\geq1-E^{N+1}\exp{N}\exp{-(N+\tfrac{1}{2})\log N}\\
            %&\geq1-E^{N}\exp{N}\exp{-N\log N}\\
            &\geq1-\exp{N(1+{\log E}) -N\log N}.
    \end{align}
    There is an $N(E)\in\mathbb N$ such that $\log N(E)\geq 1 + (1+{\log E})$, and thus for all $N\in\mathbb N$ satisfying $N\geq N(E)$ we have
    \begin{align}
        tr(P_N|\alpha\rangle\langle\alpha|)\geq1-\exp{-N}.
    \end{align}
\end{remark}
Lemma \ref{lem:effectively-finite-dimensionanl-projection}, Lemma \ref{lem:energy-bound-for-gaussian-noise} and Remark \ref{rem:application-of-stirling}, can be combined to give a formula of the effective dimension for each of the three channels in Theorems \ref{thm:unknown-noise}, \ref{thm:unknown-phase}, \ref{thm:unknown-attenuation}: For every $\alpha$ with $|\alpha|^2\leq E$ we get, with $E'>0$ and $N\geq 2+\log( E + E' )$, 
    \begin{align}
        tr&(P_N\mathcal N_\sigma(\alpha))\nonumber\\
            = &tr\left(P_N \frac{1}{\sigma\pi}\int_{C_{E + E'}}\exp{-\tfrac{|\alpha-z|^2}{\sigma}}|z\rangle\langle z|dz\right)\\
                \qquad+ & \frac{1}{\sigma\pi}tr\left(P_N \int_{C_{E + E'}^\complement}\exp{-\tfrac{|\alpha-z|^2}{\sigma}}|z\rangle\langle z|dz\right)\\
            \geq &\tfrac{1}{\sigma\pi}\int_{C_{E + E'}^\complement}\exp{-\tfrac{|\alpha-z|^2}{\sigma}}dz\cdot\min_{z\in C_{E + E'}^\complement}\langle z,P_N z\rangle\\
            \geq &\left(1-\exp{-\frac{E'}{\sigma}}\right)\left(1-\exp{-N}\right).
    \end{align}
If we choose $E'(N)=\sigma N$ then there is an $N'(E)$ such that for all $N\geq N'(E)$ we get
    \begin{align}
        tr(P_N\mathcal N_\sigma(\alpha))
            &\geq1-2\exp{-N}.
    \end{align}
    If we consider a block-length of $k$ and let $N(k)=c_d\log(k)$ for some $c_d>0$ we therefore get $tr(P_{N(k)}\mathcal N_s(\alpha))\geq1-2\cdot k^{-c_d}$, uniformly for all $\alpha$ satisfying $|\alpha|^2\leq E$. 
    %{\color{red} We should be careful with consistency of the various constraints here: $E, E'$ are fixed by the problem and we only have the freedom to choose $N$ properly. I think it's ok since we are asking $N\geq 2+\log(E+\sigma N)$.}
    
\begin{remark}\label{rem:worst-case-tails}
    While these estimates are of a simple form, they already cover a large class of channels of practical interest. Interestingly, they are far from the expected worst-case behaviour, which can be estimated as follows: Let $\rho$ be diagonal in the number state basis, with eigenvalues $\lambda_n=c/n^{-3}$ for some suitable $c>0$. Then the energy of $\rho$ for $\mathbb{H}=\sum_i i|i\rangle\langle i|$ is $tr(\rho \mathbb{H})=\sum_ic\cdot i^{-2}<\infty$. However, $\sum_{n=N}^\infty c/n^{-3}$ scales approximately as $\mathcal O(N^{-2})$, thus we only get $tr(P_N\rho)\geq1 - 1/N^{2}$, an accuracy of approximation that is not sufficient for our techniques.
\end{remark}
%%%%%%%%%%%%%%%%%%%%%%%%%%%%%%%%%%%%%%%%%%%%%%%%%%%%%%%%%%%%%%%%%%%%%%%%%
    To derive Theorem \ref{thm:unknown-noise} from Theorem \ref{thm:compound-coding-theorem} we require the following additional information: The set $\mathcal R:=\{\alpha:|\alpha|^2\leq E\}$ is closed and convex. Each state $\alpha$ has expected energy $\langle \alpha,\mathbb{H}\alpha\rangle=|\alpha|^2$. The expected output energy of a channel $\mathcal N_\sigma$ is therefore
    \begin{align}
        tr(\mathcal N_\sigma(\alpha) \mathbb{H})
            %&=\frac{1}{\sigma\pi}\int\exp{-\frac{|z-\alpha|^2}{\sigma}}|z|^2dz\\
            %&=\frac{1}{\sigma\pi}\pi \sigma(\sigma+|\alpha|^2)\\
            &=\sigma+|\alpha|^2\\
            &\leq\sigma+E.
    \end{align}
    Thus $E_\mathrm{out}=E+B$ for this channel. The optimal input distribution for the Gaussian channel is independent of $\sigma$ (see \cite[Equation (91)]{holevo1979} and therefore, since the capacity of the Gaussian channel is monotonically decreasing with $\sigma$, we get
    \begin{align}
        C(\mathcal N)\geq g(B+E)-g(B).
    \end{align}
    Obviously the reverse inequality holds as well, so that Theorem \ref{thm:unknown-noise} is proven.
%%%%%%%%%%%%%%%%%%%%%%%%%%%%%%%%%%%%%%%%%%%%%%%%%%%%%%%%%%%%%%%%%%%%%%%%%
    To derive Theorem \ref{thm:unknown-phase} from Theorem \ref{thm:compound-coding-theorem} we note that the map $|\alpha\rangle \to |\exp{\mathbbm i\theta}\alpha\rangle$ is unitary. Thus for each $\theta$, choosing the optimal distribution \cite[Equation (91)]{holevo1979}, yields a capacity 
    \begin{align}
        C(\mathcal N)= g(\sigma +E)-g(\sigma ).
    \end{align}
    Since the optimal distribution does not depend on $\theta$, the compound channel capacity of $\mathcal N$ equals 
    $g(\sigma + E) - g(\sigma)$.
%%%%%%%%%%%%%%%%%%%%%%%%%%%%%%%%%%%%%%%%%%%%%%%%%%%%%%%%%%%%%%%%%%%%%%%%%
    To derive Theorem \ref{thm:unknown-attenuation} from Theorem \ref{thm:compound-coding-theorem} we choose again the optimal distribution \cite[Equation (91)]{holevo1979} for the Gaussian channel. Energy bounds carry over as well. For every single attenuation channel, using this distribution effectively translates the problem into a transmission under energy constraint $\eta E$ so that one can show
    \begin{align}
        C(\mathcal N_\eta)\geq g(\sigma +\eta E) - g(\sigma).
    \end{align}
    Since $x\to g(x)$ is monotonously increasing (see e.g. \cite[Equation (85)]{holevo1979}) we see that 
    \begin{align}
        C(\mathcal N) = g(\sigma + A\cdot E) - g(\sigma).
    \end{align}
   The same input distribution is optimal for any pure attenuation channel~\cite{Giovannetti2004}, so that the results of Theorem \ref{thm:unknown-attenuation} apply also to the case $\sigma=0$.
%%%%%%%%%%%%%%%%%%%%%%%%%%%%%%%%%%%%%%%%%%%%%%%%%%%%%%%%%%%%%%%%%%%%%%%%%
    
%%%%%%%%%%%%%%%%%%%%%%%%%%%%%%%%%%%%%%%%%%%%%%%%%%%%%%%%%%%%%%%%%%%%%%%%%

%%%%%%%%%%%%%%%%%%%%%%%%%%%%%%%%%%%%%%%%%%%%%%%%%%%%%%%%%%%%%%%%%%%%%%%%%
%%%%%%%%%%%%%%%%%%%%%%%%%%%%%%%%%%%%%%%%%%%%%%%%%%%%%%%%%%%%%%%%%%%%%%%%%
\section{Proofs}
\begin{IEEEproof}[Proof of Lemma \ref{lem:energy-bound-for-gaussian-noise}]
    \begin{align}
        \frac{\int_{C_{E+E'}}\exp{-\frac{|\alpha-z|^2}{\sigma}}dz}{\pi \sigma}
            &\leq\frac{1}{\pi\sigma}\int_{C_{E'}}\exp{-\frac{|z|^2}{\sigma}}dz\\
            &=\frac{1}{\pi\sigma}\int_{r^2>E'}\exp{-\frac{r^2}{\sigma}}r d\theta dr\\
            &=\frac{1}{\sigma}\int_{E'}^\infty e^{-\frac x\sigma}dx=e^{-E'/\sigma}.
    \end{align}
\end{IEEEproof}

\begin{IEEEproof}[Proof of Lemma \ref{lem:effectively-finite-dimensionanl-projection}]
    Let $|n\rangle$ be the photon-number states. Then any coherent state can be written as $|\alpha\rangle = \exp(-|\alpha|^2/2)\sum_{n=0}^\infty\frac{\alpha^n}{\sqrt{n!}}|n\rangle$. Define $P_N:=\sum_{n=0}^{ N-1}|n\rangle\langle n|$, then if $|\alpha|^2\leq\frac{ N+1}{2}$ we have 
    \begin{align}
        tr(P_N|\alpha\rangle\langle\alpha|)
            %&=\sum_{n=0}^{N-1}|\langle\alpha,n\rangle|^2\\
            %&=\exp(-|\alpha|^2)\sum_{n=0}^{N-1}|\frac{(|\alpha|^2)^n}{n!}|^2\\
            %&=\exp(-|\alpha|^2)\left(\exp(|\alpha|^2)-\sum_{n=N}^\infty\left|\frac{(|\alpha|^2)^n}{n!}\right|^2\right)\\
            %&\geq\exp(-|\alpha|^2)(\exp(|\alpha|^2) -  2\frac{|\alpha|^{N}}{N!})\\
            % find this estimate in "exponentialRemainderEstimate.png"
            %&=1-2\exp(-|\alpha|^2)2\frac{|\alpha|^{N}}{N!}).
            &=1-\exp(-|\alpha|^2)\sum_{n=N}^\infty\frac{(|\alpha|^2)^n}{n!}\\
            &\geq1-\exp(-|\alpha|^2)2\frac{|\alpha|^{2N}}{N!}.
    \end{align}
    Thus the inequality is proven. The equality follows by definition of $P_N$.
\end{IEEEproof}

\begin{IEEEproof}[Direct Part of Theorem \ref{thm:compound-coding-theorem}]
%%%%%%%%%%%%%%%%%%%%%%%%%%%%%%%%%%%%%%%%%%%%%%%%%%%%%%%%%%%%%%%%%%%%%%%%%
Let $\eps>0$, $k\in\mathbb N$ and $\efd(\eps)$ be the effective dimension of $\mathcal H$. Define for each $s\in\mathbf S$
\begin{align}\label{def:finite-dim-approximation}
    \mathcal N_{s,\eps}(\alpha):=P_\eps\mathcal N_s(\alpha)P_\eps + \tfrac{tr(P_\eps^\perp\mathcal N_s(\alpha))}{tr(P_\eps)}P_\eps
\end{align} 
and let $\mathcal C_k$ be a $(k,\lambda)$ code for $\mathcal {\cal N}_{s,\epsilon}$. Define $r(\alpha):=tr(P_\eps^\perp\mathcal N_s(\alpha))$. By assumption, $r(\alpha)\leq\eps$. Setting $\mathcal N_{s,\eps}^\perp(x):=\tfrac{tr(P_\eps^\perp\mathcal N_s(\alpha))}{tr(P_\eps)}P_\eps$ we'll derive a bound on the error of this code when used for $\mathcal N$ instead as follows:
\begin{align}
    p_{\mathrm{success}}&(\mathcal C_k)\geq
    %\frac{1}{M}\sum_{m=1}^Mtr(\mathcal N_s^{\otimes k}(x^k_m)D_{m})\\
        %&=\frac{1}{M}\sum_{m=1}^Mtr(P_\eps^{\otimes k}\mathcal N_s^{\otimes k}(x^k_m)P_\eps^{\otimes k} D_{m})\\
        \frac{1}{M}\sum_{m=1}^Mtr((\mathcal N_{s,\eps}-\mathcal N_{s,\eps}^\perp)^{\otimes k}(x^k_m) D_{m})\\
        &\geq\sum_{m=1}^M\frac{tr(\mathcal N_{s,\eps}^{\otimes k}(x^k_m)D_{m})}{M} -\eps\\
        &\geq1-\lambda_k -\eps.
    \end{align}
    %{\color{red} or define for ${\cal N}_s$ the POVM $\{D'_m=P_\epsilon D_m P_\epsilon\}_{m=1}^M$ and we have
    %\begin{align}
    %p_s&(\mathcal C_k)=\frac{1}{M}\sum_{m=1}^M tr({\mathcal N}_s^{\otimes k}(x^k_m)D'_{m})\\
    %    &=\frac{1}{M}\sum_{m=1}^M tr\Big[{\mathcal N_{s,\eps}}^{\otimes k}(x^k_m)D_{m} \\
    %    &-({\mathcal N_{s,\eps}}^{\otimes k}(x^k_m)-P_\epsilon^{\otimes k} {\mathcal N}_s^{\otimes k}(x^k_m) %P_\epsilon^{\otimes k})D_m\Big]\\
    %    &\geq 1-\lambda_k - \frac{1}{M}\sum_{m=1}^M (1-tr(P_\epsilon^{\otimes k}{\mathcal N}_s^{\otimes %k}(x^k_m)))\\
    %    &\geq 1-\lambda_k - \epsilon.
    %\end{align}
    %}
    Thus, if $(\eps_k)_{k\in\mathbb N}$ satisfies $\lim_{k\to\infty}\eps_k=0$ and $(\mathcal C_k)_{k\in\mathbb N}$ is a sequence of codes - where each $\mathcal C_k$ is a $(k,\lambda_k)$ code for $\mathcal N_{s,\eps_k}$ - the sequence is automatically a sequence of codes for $\mathcal N_s$, at the same rate. 
    In the remainder of this proof, the dependence of $\lambda_k$ on $d$ will be of vital importance. Let us consider the random code as described in \cite{bbjn2012}. It holds
    \begin{lemma}[{\cite[Lemma 1]{bbjn2012}}]
        Let $\{\mathcal N_s\}_{s\in\mathbf S}$ be a compound channel and $p\in\mathcal P(\mathbf X)$. Define $\mathbf{p}:=\sum_xp(x)|e_x\rangle\langle e_x|$, 
        \begin{align}
            \rho_{k}&:=\frac{1}{|\mathbf S|}\sum_s\sum_{x^k}p^{\otimes k}(x^k)|e_{x^k}\rangle\langle e_{x^k}|\otimes \mathcal N_s^{\otimes k}(x^k)\\
            \sigma_{k}&:=\frac{1}{|\mathbf S|}\sum_s\mathbf{p}^{\otimes k}\otimes \sum_{x^k}p^{\otimes k}(x^k)\mathcal N_s^{\otimes k}(x^k).
        \end{align}
        If there is a projector $q_k$ such that
        \begin{align}
            tr(q_k\rho_{k})&\geq1-\lambda,\qquad
            tr(q_k\sigma_{k})\leq2^{-k\cdot a}
        \end{align}
        then for any $\gamma$, with $0<\gamma\leq a$, there is a code with $M=\lceil2^{k(a-\gamma)}\rceil$ and
        \begin{align}
            \frac{1}{M}\sum_{m=1}^{M}tr(\mathcal N_s(x^k_m)(\eins - D_{m}))\leq|\mathbf S|(2\cdot \lambda + 4\cdot2^{-{k}\gamma})
        \end{align}
    \end{lemma}
    \begin{lemma}
        For every $\delta>0$ and $p\in\mathcal P(\mathcal X)$ there is a $\tilde c$ such that, for every large enough $k$, there is a projector $q_k$ satisfying
        \begin{align}
            tr(q_k\rho_{k})&\geq1-|\mathbf S|\cdot 2^{-k\cdot\tilde c},\ \ \ tr(q_k\sigma_{k})&\leq2^{-k\cdot(a-\delta)},
        \end{align}
        where $a:=\min_{s\in\mathbf S}D(\rho_{s,1}\|\mathbf p\otimes \sigma_{s,1}) = \min_{s\in\mathbf S}\chi(p,\mathcal N_s) $.
    \end{lemma}
        Critical parameters of the proof in \cite{bbjn2012} are the $w(k)$, where $w(k) := \tfrac{d^2}{k}\log(k + 1)$, as introduced in \cite[(63)]{bbjn2012}, has now an additional dependency on $k$ through $d=\efd(\eps)$. The estimate \cite[(74)]{bbjn2012} translates to our setting as
        \begin{align}
            f'_{k,\nu_k}(0)\leq-\tfrac{\delta}{2}+\tfrac{1}{k}\log|\mathbf S|
        \end{align}
        and is valid as long as 
        %$0 < \nu < \nu_0(\delta)$ , where $\nu_0(\delta)$ is such that
        $\nu_k$ satisfies $2\nu_k\log\tfrac{\efd{\eps}}{2\nu_k}<\delta/2$ (see \cite{bbjn2012}, below (74)). Choosing $\nu_k=\tfrac{1}{2}2^{-c_1\cdot k}$ for arbitrary $c_1>0$ the latter inequality transforms to 
        \begin{align}
            2^{-c_1\cdot k}(\log(\efd(\eps)) + c_1\cdot k)<\tfrac{\delta}{2},
        \end{align}
        which holds true whenever $\efd(\eps)$ scales slow enough with $k$ (as in the requirement of Theorem \ref{thm:compound-coding-theorem}) and $k$ is chosen large enough.
        
        Thus there is an $s'>0$ and a $k_1=k_1(\delta,|\mathbf S|)\in\mathbb N$ such that $f_{k,\nu_k}(s)<0$ for all $s\in(0,s')$. Letting $s'$ be the number achieving $\min_sf_{k,\nu_k}(s)$ we get, with $c_2:=-f_{k,\nu_k}(s')>0$, the estimate 
        \begin{align}
            tr(q_k\rho_k)\geq1-\exp{-k(c_2 + w(k))}
        \end{align}
        Thus whenever $\lim_{k\to\infty}w(k)=0$ holds, we have proven a direct coding theorem. 
        %In the particular case treated here, we can use an effective dimension $\efd(2\cdot k^{-c_d})=c_d\log(k)$, so that making this dependence explicit we need to write $w(k) = \tfrac{c_d^2\log(k + 1)^3}{k}$. It holds $\log(k+1)=\log(k) + \log(1+1/k)$ and thus for large enough $k$ we can use $w(k) = \tfrac{(c_d\log k)^4}{k}$ in our calculations. For this choice, it still holds $\lim_{k\to\infty}w(k)=0$.
        
        Thus, $\lim_{k\to\infty}\max_{p\in\mathcal P(\mathbf X)}\min_{s\in\mathbf S}\chi(p;\mathcal N_{s,\eps_k})$ 
        is achievable under our assumptions.
        %for all sequences $\eps_k=k^{-d_c}$ with $d_c>1$. 
        It remains to show that this value converges to the proposed one for $\eps_k\to0$. Our argument rests on the continuity of entropy on the sets $\mathcal P_{\mathbb{H},E}$ \cite{wehrl1978} in the concrete form given in Lemma \ref{lem:gibbs-hypothesis}.
        This result was already used successfully for proving coding theorems in \cite{SHIROKOV201881,shirokov-uniform-approximation}. The bound on $|S(\rho) - S(\sigma)|$ does not depend on $\rho$ or $\sigma$ explicitly. All signal states obey $tr(\mathcal N_s(x) \mathbb{H})\leq E_\mathrm{out}$ by assumption. 
        %In the Gaussian case, our earlier analysis shows that all states at the output of the channel obey the inequality $tr\rho H\leq E'(E,\sigma)$ if $|\alpha|^2\leq E$. 
        For every $\rho$, its modified finite-dimensional approximation $\rho_\eps:=P_\eps\rho P_\eps + (1-tr(P_\eps\rho))\pi_\eps$ obviously satisfies 
        \begin{align}
            tr(\mathbb H\rho_\eps)\leq tr(\mathbb H\rho)\leq E_\mathrm{out}.
        \end{align}
        %Intuitively, this is because our approximation maps high energy states to low-energy ones. 
        Thus if $\lim_{\eps\to0}\|\rho_\eps-\rho\|_1=0$ then also $\lim_{\eps\to0}S(\rho_\eps)=S(\rho)$. That $\|\rho_\eps-\rho\|_1\to0$ follows from 
        %\begin{align}
        %    \|P_\eps^\perp\rho P_\eps\|_1\leq\sqrt{tr(P_\eps^\perp\rho)}
        %\end{align}
        %(see \cite[Poof of Lemma 11.1]{holevo-book} and \cite[equation (2)]{shirokov-uniform-approximation}) as follows: \red{check again}
        %\begin{align}
        %    \|\rho_\eps-\rho\|_1
            %\|&\rho_\eps-\rho\|_1\\
                %&=\|\eps'P_\eps\rho P_\eps + \eps P_\eps^\perp\rho P_\eps^\perp - (P_\eps + P_\eps^\perp)\rho(P_\eps + P_\eps^\perp)\|_1\\
                %&\leq\eps'\|P_\eps\rho P_\eps - (P_\eps + P_\eps^\perp)\rho(P_\eps + P_\eps^\perp)\|_1\\
                %&\ \ + \eps\| P_\eps^\perp\rho P_\eps^\perp - (P_\eps + P_\eps^\perp)\rho(P_\eps + P_\eps^\perp)\|_1\\
         %       &\leq\eps'\|P_\eps\rho P_\eps^\perp + P_\eps^\perp\rho P_\eps + P_\eps^\perp\rho P_\eps^\perp\|_1\\
          %      &\ \ + \eps\| P_\eps\rho P_\eps^\perp + P_\eps^\perp\rho P_\eps + P_\eps\rho P_\eps\|_1\\
                %&\leq\eps'(2\sqrt{tr(P_\eps^\perp\rho)} + \|P_\eps^\perp\rho P_\eps^\perp\|_1) + 3\eps\\
                %&\leq\eps'(2\sqrt{\eps} + \eps) + 3\eps\\
           %     &\leq6\sqrt{\eps}.
        %\end{align}
        %{\color{red} or
        %\begin{align}
        %    \|\rho_\eps-\rho\|_1 &\leq \|\frac{P_\epsilon\rho P_\epsilon}{\tr(P_\epsilon %\rho)}-\rho\|_1+\|\frac{P_\epsilon\rho P_\epsilon}{\tr(P_\epsilon \rho)}-\rho_\epsilon\|_1\\
        %    &\leq 2\sqrt{\epsilon}+\frac{1-\tr(P_\epsilon \rho)}{\tr(P_\epsilon \rho)}\|P_\epsilon \rho %P_\epsilon-P_\epsilon\|_1\\
        %    &\leq 2\sqrt\epsilon+(1-\tr(P_\epsilon \rho))\cdot 2\leq 2 (\sqrt\epsilon+\epsilon)
        %\end{align}
        %where we have used 
        the triangle inequality and the gentle measurement lemma~\cite{Winter1999}.
        %and the definition $\pi_\epsilon:=P_\epsilon/tr(P_\epsilon)$.
        %}
        Thus   
        \begin{align}
            |\chi(p;\mathcal N_{s,\eps_k}) - \chi(p;\mathcal N_{s})|\leq 2(\eps C(\eps_k,\mathbb H,E) + h(\eps_k))
        \end{align}
        for every distribution $p\in\mathcal P_f(\mathbf X)$. As a consequence, for every finite subset $\mathbf X\subset\mathcal R$ we have
        \begin{align}
            C(\mathcal N)\geq\max_{p\in\mathcal P(\mathbf X)}\min_{s\in\mathbf S}I(p;\mathcal N_{s}).
        \end{align}
        To prove the corresponding statement for general $\mathcal S$ and arbitrary $\mathbf X\subset\mathcal X$ we cover $\mathcal S$ with a discrete net which scales as $|\mathcal S_\alpha|\leq(\frac{6}{\alpha})^{2|\mathbf X|\efd(\eps_k)^2}$ and delivers, for every $s\in\mathcal S$ and $x^k\in\mathbf X^k$, an $s'\in\mathbf S_\alpha$ such that $\|\mathcal N_s(x^k)^{\otimes k} - \mathcal N_{s'}(x^k)^{\otimes k}\|_1\leq2\cdot k\cdot\alpha$ \cite[Lemma 6]{bbjn2012}. 
        We pick $\alpha_k=k^{-2}$. Then any code for the finite compound is asymptotically optimal for the infinite one as well. Moreover, in the particular case treated here,
        \begin{align}
            |\mathcal S_{\alpha_k}|\cdot 2^{-\tilde c\cdot k}\leq (6k)^{4|\mathbf X|\efd(\eps_k)^2}2^{-\tilde c\cdot k}
        \end{align}
        and thus for every $\delta>0$, finite set $\mathbf X\subset\mathcal X$ of signals and distribution $\mathbf p$ over the signals, $\lim_{k\to\infty}\inf_{s\in\mathbf S_{1/k^2}}\chi(p;\mathcal N_{s,\eps_k})-\delta$ can be achieved. By the same continuity arguments as above, this implies Theorem \ref{thm:compound-coding-theorem}. 
    \end{IEEEproof}
    
    \begin{IEEEproof}[Converse Part of Theorem \ref{thm:compound-coding-theorem}]
        If $(\mathcal C_k)_{k\in\mathbb N}$ is a sequence of codes for $\mathcal N$ achieving rate $R>0$ then the sequence $(\mathcal C'_{k})_{k\in\mathbb N}$ obtained by adjusting all POVM elements $D^{k}_m$ of $\mathcal C_k$ as $D^{(k)}_m\to P_{N(k)} D^{(k)}_m P_{N(k)}$ (where $N(k)$ is chosen such that the approximation parameter $\eps_k=k^{-2}$ in \ref{def:finite-dim-approximation}) achieves the same rate $R$ for $\{\mathcal N_{s,k^{-2}}\}_{s\in\mathcal S}$ as in \eqref{def:finite-dim-approximation}. After discrete approximation \cite[Lemma 6]{bbjn2012} of $\mathcal P(\mathcal \mathrm{supp}(P_{N(k)}))$ the converse proof of \cite{bbjn2012} applies, with $d$ replaced by $\efd(k^{-2})$ and with alphabets $\mathbf X_k$ of size $|\mathbf X_k|\in\mathcal O(k^{\efd(k^{-2})^2})$. The dependence of our approach on $\efd(k^{-2})$ can be picked up from the converse in \cite{Winter1999}. Since by assumption $\efd(k^{-2})\in\mathcal O(k^{2^{-1}+\tau})$ for some $\tau>0$, Lemma \ref{lem:gibbs-hypothesis}) lets us prove that 
        $R\leq \sup_{p\in\mathcal P_{\mathcal R}}\inf_{s\in\mathcal S}\chi(p;\mathcal N_s)$. 
    \end{IEEEproof}

    \section{Application: Kennedy Receiver Performance under Compound Loss}
Here we consider the rate attained by a simple receiver on a compound lossy channel with coherent-state input. We let $\eps\in(0,1)$ and $\mathcal N = \{{\cal N}_{\eta}\}_{\eta\in\mathbf S}$, with $\mathbf S=\{\eps,1\}$, be a compound channel consisting of pure loss channels
\begin{equation}
{\cal N}_{\eta}:\ket\alpha\mapsto\ket{\sqrt\eta \alpha}.
\end{equation} 
The Kennedy receiver~\cite{Kennedy1973} is a standard receiver for the discrimination of two coherent states $\alpha_{x}:=a(-1)^{x}$, with $a\in\mathbb{R}$ and $x=\{0,1\}$. It employs a displacement operation $\ket\alpha\mapsto\ket{\alpha-\beta}$, where $\beta=b\in\mathbb{R}$ if $\alpha\in\mathbb{R}$ without loss of generality, and a threshold photodetector, represented by a quantum measurement $\{\ket{0}\bra{0},\mathbf{1}-\ket{0}\bra{0}\}$. This receiver, optimized over $b$, beats the homodyne receiver for $a\gtrsim0.2$~\cite{Weedbrook2012} and has an adaptive refinement, the Dolinar receiver~\cite{Dolinar1973}, which asymptotically attains the minimum error probability for discrimination. 

We now show that naively optimizing the Kenneday receiver for the worst channel ($\mathcal N_\eps$ in this case) is not optimal. We employ the binary alphabet $\{(\alpha_{x};p_{x})\}_{x=0,1}$ at the sender side to communicate over ${\cal N}$ and a Kennedy receiver with displacement $b$ at the receiver side. The induced classical channel has output $y\in\{0,1\}$ and transition function defined by
\begin{equation}
p_{\eta,b}(0|x)=e^{-(b-\sqrt\eta \alpha_{x})^{2}}, p_{\eta,b}(1|x)=1-p_{\eta,b}(0|x).
\end{equation}
Our strategy of proof is to send signals at high energy, such that we become able to produce analytical estimates on the capacity of $p_{\eta,b}$.
\begin{figure}
\includegraphics[width=0.5\textwidth]{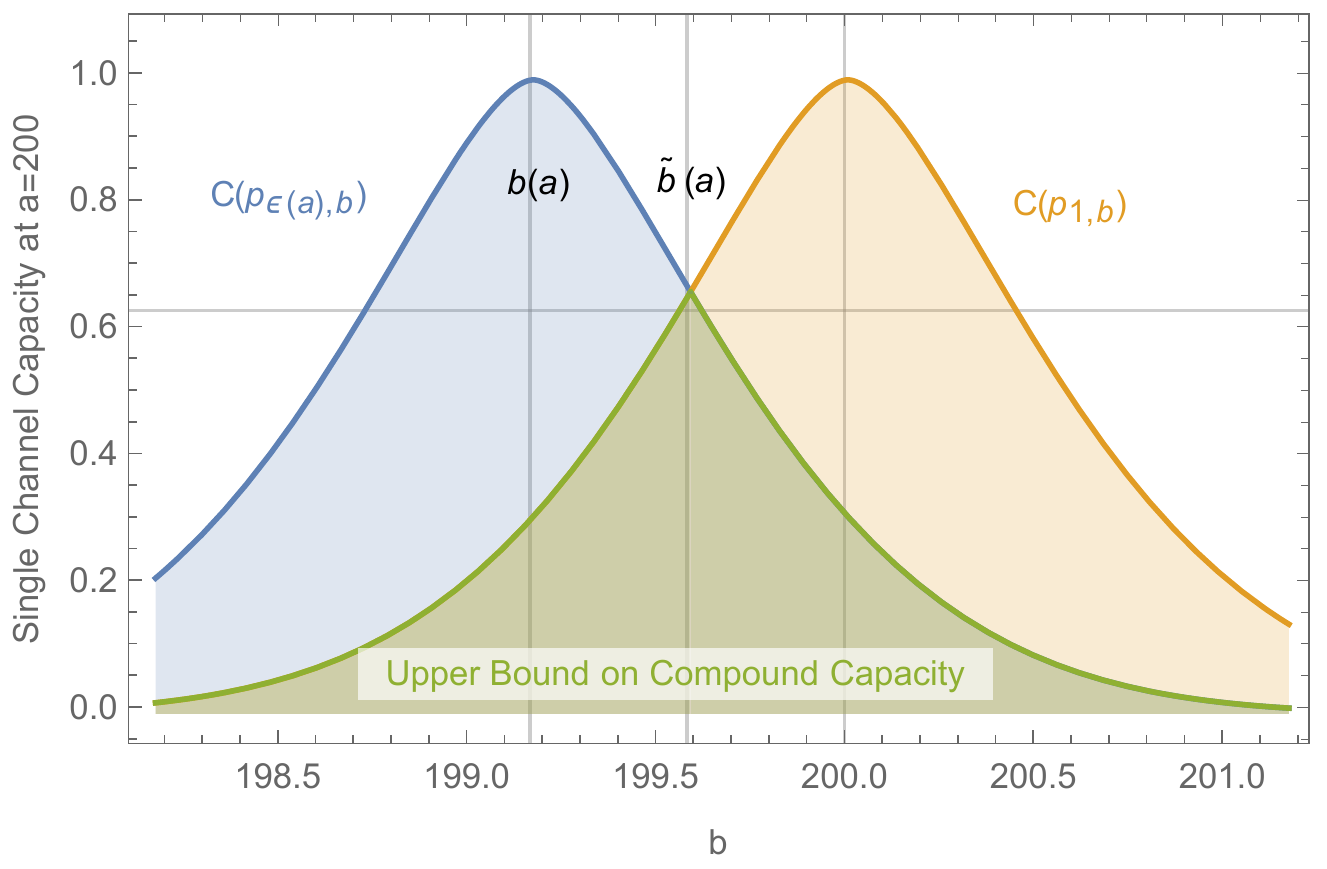}
\caption{Figure 1: Numerical Estimates on Single Channel Capacities and on Compound Channel Capacity for $a=200$ and $\eps=\eps(a)\approx0.996$.}
\end{figure}
Let $\delta>0$, $a\in\mathbb R$ and $\eps\in(0,1)$. Set $b=\sqrt{\eps}a$. We explain below how this choice of $b$ is both almost-optimal for $\mathcal N_\eps$ at high power levels $a$ and yet highly non-optimal for the compound channel $\mathcal N$. With our choice of $b$ it holds
\begin{align}
    p_{\eps,b}(0|0)=1,\qquad\qquad\qquad &p_{\eps,b}(0|1)=e^{-4\eps\cdot a^2}\\
    %\\
    %p_{\eps,b}(0|1)&=e^{-(\sqrt{\eps}a + \sqrt{\eps}a)^{2}}=e^{-4\eps\cdot a^2}.
    p_{1,b}(0|0)=e^{-a^2(\sqrt{\eps} - 1)^{2}},\qquad &p_{1,b}(0|1)=e^{-a^2(\sqrt{\eps} + 1)^{2}}.
\end{align}
Define, for every $p\in[0,1]$, $w_p\in C(\{0,1\},\{0,1\})$ by
\begin{align}
    w_p(y|0)=p\delta_0(y)+p'\delta_1(y),\qquad w_p(y|1)=\delta_1(y).
\end{align}
Let $c>0$. Choosing $\eps=\eps(a)=((a-c)/a)^2$ we get $\lim_{a\to\infty}\eps(a)=1$ and for every $a>0$ we have
$b=b(a)=\sqrt{\eps(a)}a=a-c$. It then holds uniformly for all $c>0$ that
\begin{align}
    \lim_{a\to\infty}\|p_{\eps(a),b(a)}-w_1\|_F=0.
\end{align} 
In addition,
\begin{align}
    p_{1,b}(0|0)&=e^{-c^{2}},\qquad   p_{1,b}(0|1)=e^{-(2a-c)^{2}}.
\end{align}
With the special choice $c=\sqrt{\ln 2}$ we get 
\begin{align}
    \lim_{a\to\infty}\|p_{1,b(a)}-w_{2^{-1}}\|_F=0.
\end{align}
The compound channel capacity $C$ \cite{bbt59} is continuous. Therefore, 
\begin{align}
    C(\{p_{1,b(a)},p_{\eps(a),b(a)}\})\leq C(\{w_{2^{-1}},w_1\})+\delta
\end{align} 
for large enough $a$. The channel $w_{2^{-1}}$ has capacity $\log5/4$, therefore $C(\{w_{2^{-1}},w_1\})\leq\log5/4$. It follows that there exists an $a_1>0$ such that for all $a>a_1$
\begin{align}
    C(\{p_{1,b(a)},p_{\eps(a),b(a)}\})\leq \log5/4+\delta,
\end{align}
and $a_2>0$ such that for all $a>a_2$
\begin{align}
    C(\{p_{\eps(a),b(a)}\})\geq C(\{w_1\})-\delta=1-\delta.
\end{align}
Thus we can state: For all $a$ satisfying $a>a_1+a_2$, if we choose $b(a)$ as the Kennedy receiver parameter then $b(a)$ is almost-optimal for $p_{\eps(a),b(a)}$ but (choosing $\delta$ small enough) leads to a capacity $<1/3$ for transmission over $\{p_{1,b(a)},p_{\eps(a),b(a)}\}$. 

Let us consider another choice for $b$ instead: Set $\tilde b(a)=( (a-c) + a)/2=a-c/2$ then we get
\begin{align}
    &p_{\eps(a),\tilde b(a)}(0|0)=e^{-\tfrac{c^2}{4}},\ \ 
        p_{\eps(a),\tilde b(a)}(0|1)=e^{-(2a-\tfrac{3c}{2})^2}\\
    &p_{1,\tilde b(a)}(0|0)=e^{-(\tilde b(a) - a)^{2}}, \ \ 
        p_{1,\tilde b(a)}(0|1)=e^{-(2a-\tfrac{c}{2})^{2}}.
\end{align}
Therefore
\begin{align}
    \lim_{a\to\infty}\|p_{\eps(a),\tilde b(a)}-w_{e^{-c^2/4}}\|_F=0,\\
    \lim_{a\to\infty}\|p_{1,\tilde b(a)}-w_{e^{-c^2/4}}\|_F=0.
\end{align} 
Thus for every $\delta>0$ there is an $a_3>0$ such that for all $a>a_3$ 
\begin{align}
    C(\{p_{\eps(a),b(a)}\})&\geq 1-\delta\\
    C(\{p_{1,b(a)},p_{\eps(a),b(a)}\})&\leq 3^{-1}+\delta\\
    C(\{p_{1,\tilde b(a)},p_{\eps(a),\tilde b(a)}\})&\geq C(w_{e^{-\ln(2)/4}})-\delta.
\end{align}
Since $C(w_{e^{-\ln(2)/4}})\geq 1-h(e^{-\ln(2)/4})/2\geq 1-0.75\cdot 0.5\geq0.625$ we have shown the existence of compound channels $\{p_{1,b(a)},p_{\eps(a),b(a)}\}$ with choices for $b(a)$ which are almost-optimal for $\{p_{\eps(a),b(a)}\}$ but perform strictly below optimal for $\{p_{1,b(a)},p_{\eps(a),b(a)}\}$.

    \begin{section}{Conclusion}
        We have derived capacity formulas for classical-quantum compound channels of practical interest. Furthermore, we demonstrated a nontrivial choice of the displacement parameter of the Kennedy receiver when applied to a compound channel, therewith proposing the compound channel model as a tool for receiver design in applications under timing constraints, where adaptive adjustment of the receiver is not desirable.
    \end{section}

\subsection{Acknowledgements} This work was financed by the DFG via grant NO 1129/2-1 (JN) and by the BMBF via grant 16KIS0948 (AC). This project has received funding from the European Union's Horizon 2020 research and innovation programme under the Marie Sklodowska-Curie grant agreement No 845255 (MR). 

\bibliographystyle{plain}
\bibliography{bib}
\end{document}